\documentclass[conference, 10pt]{IEEEtran}
\usepackage{amssymb}
\usepackage{amsmath, tabularx}
\usepackage{mathtools}
\usepackage{ellipsis}
\usepackage{textcomp}
\usepackage{breqn}
\usepackage{cite}
\usepackage{graphicx}	
\usepackage{subfigure}
\usepackage{epsfig}
\usepackage{psfrag}
\usepackage{grffile}

\usepackage{setspace}
\usepackage{epstopdf} 
\usepackage{color}
\usepackage{multirow}
\usepackage{amsthm}

\usepackage[overload]{empheq}
\usepackage{cases}
\usepackage{url}

\usepackage{listings}
\usepackage{xcolor}

\definecolor{codebg}{RGB}{245,245,245}
\definecolor{codegray}{RGB}{110,110,110}
\definecolor{codeblue}{RGB}{0,92,184}
\definecolor{codegreen}{RGB}{0,128,0}
\definecolor{codered}{RGB}{220,0,0}

\usepackage{pifont}
\usepackage{xcolor}
\usepackage[noend]{algpseudocode}
\usepackage{algorithm}
\algnewcommand{\LineComment}[1]{\Statex \(\triangleright\) #1}
\algnewcommand{\And}{\textbf{and}}
\algnewcommand{\Or}{\textbf{or}}
\usepackage{multicol}

\algnewcommand{\LineCommentCont}[1]{\Statex \hskip\ALG@thistlm%
	\parbox[t]{\dimexpr\linewidth-\ALG@thistlm}{\hangindent=\trianglerightwidth \hangafter=1 \strut$\triangleright$ #1\strut}}

\DeclareTextFontCommand{\textOF}{\fontfamily{lmtt}\selectfont}

\IEEEoverridecommandlockouts

\title{Vulnerability Analysis of eBPF-enabled Containerized Deployments of 5G Core Networks}
\author{\IEEEauthorblockN{Yash Deshpande and Samaresh Bera, \IEEEmembership{Senior Member, IEEE}}
	\IEEEauthorblockA{Department of Computer Science and Engineering\\
		Indian Institute of Technology Jammu\\
		Jammu and Kashmir, India - 181221\\
		Email: 2024pcs0044@iitjammu.ac.in, s.bera.1989@ieee.org}
}

\begin{document}

	\maketitle

	\begin{abstract}
		The extended Berkeley Packet Filter (eBPF) is useful for faster packet processing and network monitoring in softwarized deployments. Similarly, softwarized deployments of 5G core network services adopted eBPF to meet the stringent latency and bandwidth requirements of underlying applications. While the existing studies focused on network performance, security concerns over eBPF-enabled platforms are overlooked.
		
		In this paper, we study the vulnerability analysis of 5G core network deployments that use eBPF for packet processing and traffic monitoring. In particular, we consider the following aspects: a) tracing, b) denial-of-service (DoS), c) stealing information, and d) bash injection. We present the detailed attack scenarios with step-by-step implementation of containerized and eBPF-enabled 5G network functions using Open5GS. The experiment results show that the aforementioned vulnerabilities are present in eBPF-enabled 5G deployments and can be exploited by attackers. Finally, we present some mitigation techniques useful for addressing the vulnerabilities. The source code and implementation details are made available at \url{https://github.com/chimms1/5G-eBPF-exploits}.
	\end{abstract}


	\section{Introduction}\label{Sec:Introduction}
	The service-based architecture of 5G network allows the network operators to place 5G core network functions (NFs) on cloud platforms in a containerized or virtual machine form~\cite{StudyManagementCloudnative2022}. Consequently, telecom operators started deploying 5G network functions on public clouds, private clouds, or hybrid clouds. For example, Nokia and Telefónica deployed a 5G Standalone (SA) core on AWS for ultra-low latency services~\cite{O2TelefonicaNokia}. Similarly, Ericsson and Swisscom ran a 5G core proof-of-concept on AWS to explore hybrid-cloud use cases~\cite{SwisscomEricssonAWS}. These deployments often use orchestration platforms (e.g., Kubernetes) to manage the containers.
	
	The applications supported by 5G and beyond networks are broadly categorized as enhanced mobile broadband (eMBB), ultra-reliable and low-latency communications (uRLLC), and massive machine-type communications (mMTC). The quality-of-service (QoS) requirements of these applications range from high bandwidth to high-reliability and low-latency~\cite{3gpp5GStudyScenarios2017}. To meet these requirements, 5G networks must minimize processing overhead and quickly route user data. Thus, optimizing the data-plane performance of 5G NFs is critical. However, softwarized deployments of the network functions may not be suitable to meet stringent QoS requirements, unlike the hardware-based deployments. Thanks to the extended Berkeley Packet Filter (eBPF)~\cite{ebpffoundationEBPFDocumentationWhat2024}, which is a promising technology for faster packet processing and network visibility. eBPF allows custom bytecode programs to run in the Linux kernel, enabling fast packet processing and in-kernel telemetry~\cite{ebpffoundationEBPFDocumentationWhat2024}. In particular, eBPF programs attached via the eXpress Data Path (XDP) mechanism can process packets at the network driver level with minimal overhead. Recent works have utilized eBPF and XDP to accelerate 5G user-plane functions~\cite{zhouCableFrameworkAccelerating2023a, scheichEXpressDataPath2023, huangAccelerating5GServiceBased2024a}. For example, Zhou \emph{et al.}~\cite{zhouCableFrameworkAccelerating2023a} showed that incorporating eBPF/XDP into a 5G UPF can reduce per-packet processing time by over 30\%. On the other hand, Scheich \emph{et al.}~\cite{scheichEXpressDataPath2023} proposed XDP extensions for high-capacity 5G UPFs. Xuan \emph{et al.}~\cite{huangAccelerating5GServiceBased2024a} presented an eBPF/XDP-enhanced 5G Service-Based Architecture (SBA) platform and reported significant gains in throughput and latency reduction. These works illustrate that container-based 5G deployments can integrate eBPF/XDP for both performance and observability.
	
    eBPF is also widely used for security and monitoring in cloud-native environments. For example, security tools like Falco~\cite{FalcoCloudNativeRuntime2024} and network tools like Cilium~\cite{isovalentCiliumEBPFbasedNetworking2024} leverage eBPF tracing and maps to inspect network flows and system calls. eBPF can attach to kernel tracepoints, kprobes, or network interfaces to collect event logs without modifying the application code. Yang \emph{et al.}~\cite{yangZeroTracerInBandEBPFBased2025} described an in-kernel tracing system (ZeroTracer) that uses eBPF to track HTTP requests across microservices with high accuracy, demonstrating eBPF's usefulness for observability. Similarly, Soldani \emph{et al.}~\cite{soldaniEBPFNewApproach2023} demonstrated the feasibility of using eBPF for high performance monitoring and security in 5G and future 6G networks. They proposed a platform to unify observability, energy estimation, and policy enforcement in the cloud. Thus, eBPF's programmability and efficiency have led to its broad adoption for monitoring, load balancing, and security policies in containerized clusters.

	The very features that make eBPF powerful can introduce new vulnerabilities. In a cloud environment where multiple containers share the host kernel, eBPF programs loaded by one container can potentially interact with or affect other containers. Recent studies have revealed that containerized eBPF programs can break out of their namespace isolation and bypass kernel protections~\cite{heCrossContainerAttacks2023, jinEPFEvilPacket2023}. For instance, attackers can exploit specific helper functions to inspect or modify host processes, regardless of container boundaries. These capabilities enable cross-container attacks, ranging from data theft to denial-of-service, highlighting a gap in current security models where default permissions grant eBPF programs access to powerful helpers without adequate isolation.

    Given these concerns, it is crucial to analyze the security of eBPF-enabled 5G deployments. There is a lack of prior studies examining how eBPF features might be abused in a 5G containerized environment. Our research objective is to identify vulnerabilities in such deployments, evaluate exploitability, and propose defenses. Specifically, we pose the following questions: 
    \begin{itemize}
    	\item \emph{What vulnerabilities exist in containerized 5G networks that use eBPF/XDP?}
    	\item \emph{Can attackers exploit eBPF helpers or programs to attack 5G components?}
    	\item \emph{What measures can mitigate these threats?}
    \end{itemize}  

    In this paper, we deploy a fully functional eBPF-enabled 5G core network, and then combine insights from existing works and experiments to address these questions. We draw on known eBPF attack primitives and adapt them to a 5G setting. The contributions of this work include a detailed threat analysis of eBPF in a containerized 5G system and a demonstration of attack procedures. The development and source code are made available at \url{https://github.com/chimms1/5G-eBPF-exploits}. The remainder of this paper is organized as follows: Section~\ref{Sec:System_arch} details the system architecture and the role of eBPF in 5G. Section~\ref{Sec:exp_setup} defines the threat model and experimental setup. Section~\ref{Sec:vul_analysis} presents the vulnerability analysis and attack algorithms. Section~\ref{Sec:mitigations} discusses mitigations, and Section~\ref{Sec:Conclusion} concludes the paper.

	\section{System Architecture}\label{Sec:System_arch}

\begin{figure}[!ht]
    	\centering
    	\subfigure[Architecture]{
    		\includegraphics[scale=0.75]{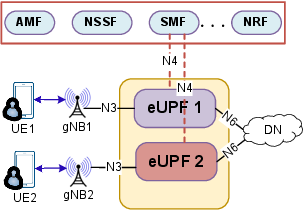}
    		\label{fig:5G_SBA}
    	}
    	\subfigure[Packet processing inside eUPF]{
    	\includegraphics[scale=0.75]{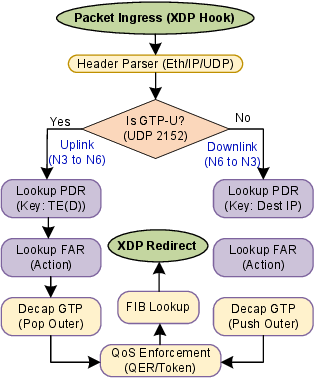}
    	\label{fig:5G_eUPF_packet_processing}
    	}
    \caption{5G service based architecture with eUPFs and XDP-based packet processing}
    \label{fig:5G_SBA_eUPF}
    \end{figure}

    We consider a 5G Standalone (SA) network architecture as defined by 3GPP, consisting of virtualized network functions (NFs), as shown in Figure~\ref{fig:5G_SBA_eUPF}. The UPF handles all user plane data forwarding between the Radio Access Network (gNB) and external Data Networks. The 5G core separates the control plane from the data plane, allowing the UPF to be scaled and placed independently.

    To simulate a multi-tenant 5G environment, our testbed deploys two distinct User Plane Function (UPF) instances. This dual-UPF configuration is intended to support network slicing, which allows network operators to create multiple virtual networks on a shared physical infrastructure. Our setup defines two distinct Network Slice Selection Assistance Information (S-NSSAI) values, with each UPF dedicated to a specific slice. This separation ensures that traffic from the User Equipment (UE) attached to different slices is routed through isolated data paths, catering to distinct use cases.

    In our deployment, each 5G NF runs as a containerized network function managed via Docker Compose. One set of containers hosts the control plane (AMF, SMF, NRF, etc.), while separate containers host the two UPF instances. The UPFs in our system are based on EdgecomLLC's open-source eUPF implementation, which utilizes eBPF/XDP for fast packet processing~\cite{edgecomllcEupfUserPlane2024}. The eUPF interacts with the 5G core via standard 3GPP interfaces: it exposes the N3 interface to receive GTP-U encapsulated packets from the Radio Access Network gNB, the N4 interface to receive Packet Forwarding Control Protocol (PFCP) rules from the SMF, and the N6 interface to route decapsulated user traffic to external Data Networks (DN), as depicted in Figure~\ref{fig:5G_SBA_eUPF}.

    Internally, the eUPF attaches an XDP program to the host's network interface to handle these flows. Incoming N3 packets are processed by this XDP program in the kernel before reaching the main userspace logic. This allows critical packet handling—such as GTP header parsing, TEID lookup, and forwarding decisions—to be executed at a faster speed. Similarly, for outgoing traffic on the N6 interface, XDP can quickly redirect packets, bypassing the substantial overhead of the Linux kernel’s standard network stack. eUPF container requires \texttt{CAP\_SYS\_ADMIN} and \texttt{CAP\_NET\_ADMIN} privileges~\cite{ebpffoundationEBPFDocumentationWhat2024} to load and attach the necessary eBPF/XDP programs.

	\section{Threat Model and Experimental Setup}\label{Sec:exp_setup}

    5G networks are supposed to be cross-vendor compatible; operators often deploy components from diverse sources~\cite{StudyManagementCloudnative2022}. This heterogeneity introduces the risk that maliciousness may originate from any part of the supply chain. In our setup, we assume that one of the eUPF containers is malicious and serves as the launchpad for exploits. This compromise can manifest in two primary ways: the container may be pre-packaged with malware (a supply chain attack), or a legitimate container may be compromised and hijacked by an attacker post-deployment. Regardless of the entry point, our analysis focuses exclusively on the host kernel's eBPF functionality as the attack vector. All NFs are deployed on the same host. Security settings or kernel configurations that are available by default are not modified in any system or container.

    Our testbed consists of the following software specifications:

    \begin{itemize}
    \item \textbf{Host OS:} Ubuntu 24.04 LTS.
    \item \textbf{Kernel:} Linux 6.14 generic.
    \item \textbf{5G Software:} Open5GS~\cite{OpenSourceImplementation} with UERANSIM~\cite{Free5GCOpenSource}.
    \item \textbf{User Plane:} EdgecomLLC eUPF (v0.6.4), an open-source high-performance UPF implementation using eBPF/XDP~\cite{edgecomllcEupfUserPlane2024}.
    \item \textbf{Container Runtime:} Docker with \texttt{docker-compose} orchestration.
\end{itemize}

 The \texttt{docker-compose} configuration grants the eUPF container specific privileges required for its operation:
\begin{verbatim}
    cap_add:
      - NET_ADMIN
      - SYS_ADMIN
    volumes:
      - /sys/fs/bpf:/sys/fs/bpf
\end{verbatim}

The complete source code for the experimental setup, attack implementations, and scripts is available at \url{https://github.com/chimms1/5G-eBPF-exploits}.

\section{Vulnerability Analysis}\label{Sec:vul_analysis}

We now analyze specific attacks enabled by eBPF in our 5G setup. Each attack exploits certain eBPF helper functions that, while useful for legitimate tasks, can be repurposed maliciously. We divide the analysis into four subsections—Tracing, Denial of Service, Information Theft, and Bash Injection—and discuss the implementation and working of each attack scenario in the subsequent sections.

\subsection{Attack Scenario 1: Tracing}

eBPF tracing enables low-overhead, programmable observability of kernel and user-space events. Few helpers, such as \texttt{bpf_get_current_comm()} and \texttt{bpf_get_current_uid_gid()} are essential for observability tools to map network traffic or \texttt{syscalls} to specific process names for debugging.

\begin{lstlisting}[language=C++,caption={Pseudocode for tracing attack},label={code:trace}]
SEC("raw_tracepoint/sys_enter")
int trace_processes(struct bpf_raw_tracepoint_args *ctx)
{
    pid = bpf_get_current_pid_tgid() >> 32;
    unsigned long long uid_gid = bpf_get_current_uid_gid();

    bpf_get_current_comm(&comm, sizeof(comm));

    /* Print or filter process info */
}
\end{lstlisting}

\begin{figure}[htbp]
	\centering
	\includegraphics[scale=0.45]{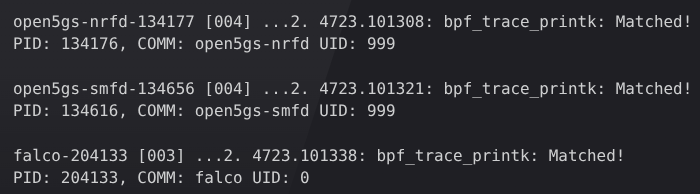}
	\caption{Output of the tracing program}
	\label{fig:outtracing}
\end{figure}

 An attacker uses these to perform reconnaissance. As shown in Listing~\ref{code:trace}, by hooking into \texttt{sys\_enter}, the attacker can enumerate all processes running on the host, identifying the PIDs of critical 5G NFs (e.g., \texttt{open5gs-amfd}, \texttt{open5gs-smfd}) or any other services that reside in separate containers. Figure~\ref{fig:outtracing} illustrates a segment of the output from the tracing program accessible to the attacker, displaying processes that are running on different containers. This breaks the process isolation guarantee of containerization.

 \subsection{Attack Scenario 2: Denial of Service (DoS)}

In general, \texttt{bpf\_send\_signal()} can be used by an eBPF program to send a signal to the current task. The attacker modifies the tracing program to send a \texttt{SIGKILL} (signal 9) to the identified security services or 5G processes. A sample attack using \texttt{bpf\_send\_signal()} is presented in Listing~\ref{code:dos}.

 \begin{lstlisting}[language=C++,caption={Pseudocode for DoS attack},label={code:dos}]
SEC("kprobe/__x64_sys_read")
int kill_falco(struct pt_regs *ctx)
{
    bpf_get_current_comm(&comm, sizeof(comm));
    if (comm == TARGET_PROCESS) {
        bpf_send_signal(9); // SIGKILL
    }
    return 0;
}
\end{lstlisting}

Since the eBPF program runs in the kernel, it bypasses container namespaces. Once the AMF or SMF process is detected, it is immediately killed. In a Kubernetes environment, the orchestrator would attempt to restart the pod, but the persistent eBPF hook would kill it again immediately, causing a \textit{CrashLoopBackOff} and rendering the 5G core offline. This is a highly effective DoS attack.

\begin{figure}[htbp]
	\centering
	\subfigure[Output of \texttt{docker ps} before commencement of attack]{
		\includegraphics[scale=0.25]{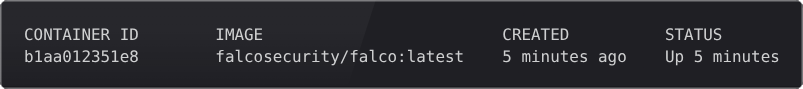}
		\label{fig:dos-a}
	}
	
	\subfigure[Attacker loads DoS program]{
		\includegraphics[scale=0.25]{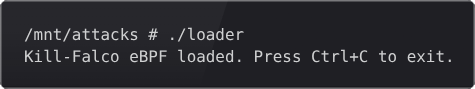}
		\label{fig:dos-b}
	}
	
	\subfigure[Output of \texttt{docker ps} after attack is loaded]{
		\includegraphics[scale=0.25]{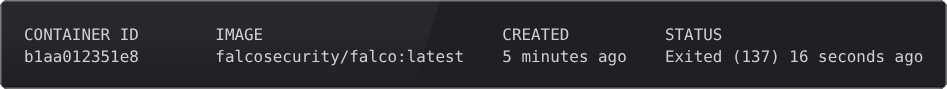}
		\label{fig:dos-c}
	}
	
	\caption{Sample execution of DoS attack}
	\label{fig:dos-output}
\end{figure}

Figure~\ref{fig:dos-output} illustrates the practical execution of this attack against a runtime security container (Falco). Initially, Figure~\ref{fig:dos-a} shows the target container in a healthy, running state. In Figure~\ref{fig:dos-b}, attacker loads the malicious eBPF object file from the compromised eUPF container. As soon as the target process invokes the hooked system call, the eBPF program triggers \texttt{bpf\_send\_signal(9)}. The result can be seen in Figure~\ref{fig:dos-c}, where the containerized process has been terminated.

\subsection{Attack Scenario 3: Information Theft}

To intercept sensitive data, the attacker must bridge the gap between kernel space execution and user space memory. This is achieved using specific eBPF helper functions: a) \texttt{bpf\_probe\_read\_user\_str()} to capture the filename arguments from the \texttt{openat} syscall, allowing the program to identify when specific sensitive files are accessed; and b) \texttt{bpf\_probe\_read\_user()} during the \texttt{read} syscall exit to extract the actual file content from the application's memory buffer. These helpers allow the eBPF program to bypass memory isolation boundaries and exfiltrate data being used by legitimate applications. The detailed implementation of the exploit can be found in the aforementioned GitHub repository.

%
%
%
%
%
%

Attacker can monitor \texttt{openat} and \texttt{read} syscalls to intercept sensitive files opened by other containers or the host. For example, stealing SSH keys or 5G encryption keys (like the \texttt{K} and \texttt{OPc} stored in UDM configurations). The attack proceeds in two phases:
\begin{itemize}
    \item \textbf{Open Detection:} When a target process (e.g., SSH or a config reader) calls \texttt{openat}, the program checks if the filename matches a sensitive target (e.g., \texttt{id\_rsa}). If so, the PID is stored in a BPF map.

    \item \textbf{Read Interception:} When the saved PID exits a \texttt{read} syscall, the eBPF program reads the content of the user-space buffer populated by the kernel and exfiltrates it.
\end{itemize}

\begin{figure}[!ht]
	\centering
	\includegraphics[scale=0.65]{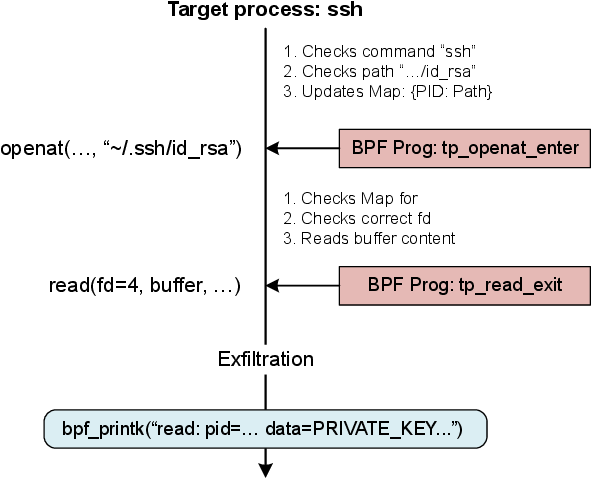}
	\caption{Working principles of information theft}
	\label{fig:information_theft_steps}
\end{figure}

\begin{figure}[htbp]
	\centering
	\includegraphics[scale=0.4]{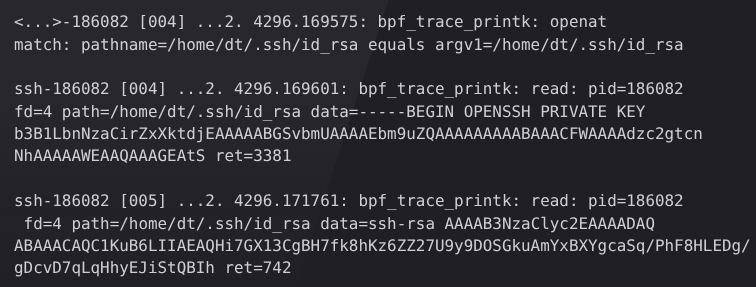}
	\caption{Output of the information theft attack, which shows compromised ssh keys}
	\label{fig:outinfotheft}
\end{figure}

Figure~\ref{fig:outinfotheft} shows an example scenario when an attacker has loaded the eBPF program from the compromised container and there is an SSH login attempt from another container. The SSH login will complete without any interruptions but on the attacker side, all the open files, including the keys, will be visible, as depicted in Figure~\ref{fig:outinfotheft}.

\subsection{Attack Scenario 4: Bash Injection}
This section details a ``Bash Injection" attack where a benign script execution is intercepted and replaced with malicious commands at runtime. While the previous attacks focused on passive monitoring, this attack causes active interference. \texttt{bpf\_probe\_write\_user()} helper allows a privileged eBPF program to overwrite data in the user-space memory of a target process. Combined with \texttt{bpf\_override\_return()}, which allows the modification of a system call's return value, an attacker can silently alter the execution flow of admin scripts. Similar to information theft, detailed implementation of the exploit can be found in the aforementioned GitHub repository.

\begin{figure}[!ht]
	\centering
	\includegraphics[scale=0.55]{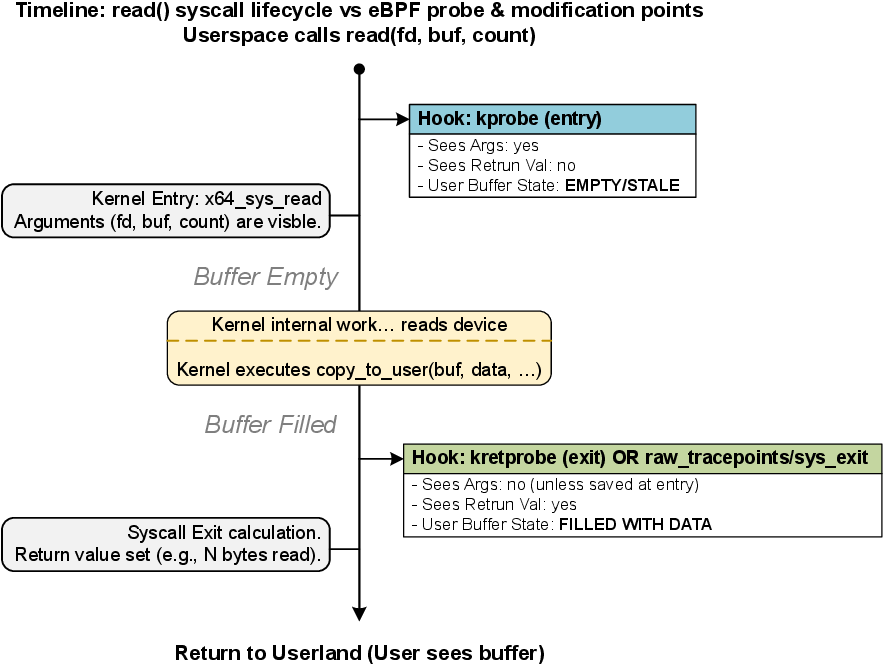}
	\caption{Lifecycle of the \texttt{read()} syscall vs. eBPF modification points}
	\label{fig:hkpts}
\end{figure}

\begin{figure}[!ht]
	\centering
	\includegraphics[scale=0.65]{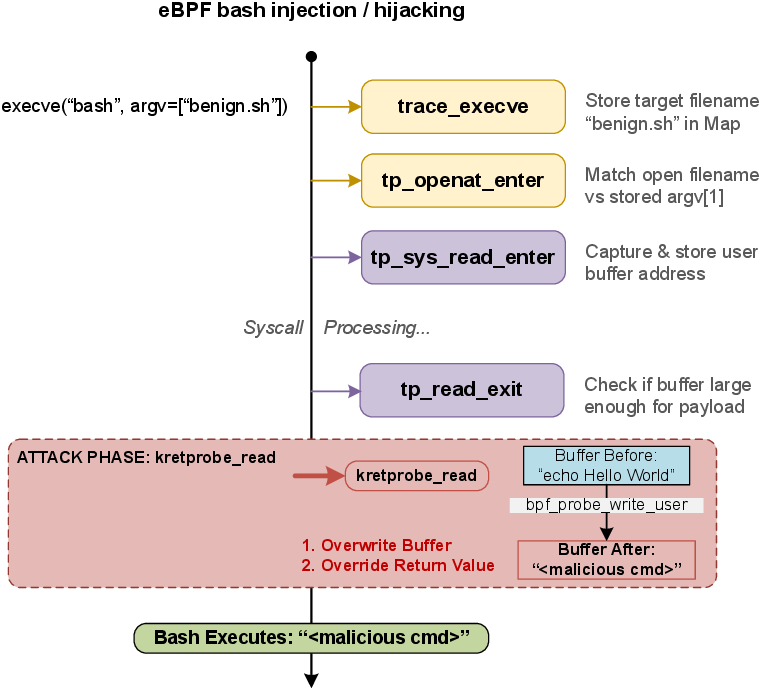}
	\caption{Steps for bash injection}
	\label{fig:bash_injection}
\end{figure}


An attacker waits for a privileged process (like a root cron job or admin bash session) to execute a benign script. The eBPF program intercepts the \texttt{read} syscall where the script content is loaded and overwrites it with malicious commands. The attack monitors \texttt{execve} to identify when a target script is launched. When the interpreter reads the script file, the eBPF hook overwrites the buffer with a malicious payload and adjusts the return value to match the new payload length, ensuring the interpreter executes the attacker's code instead of the original script. This grants the attacker root-level execution on the host.

The mechanics of this injection rely on the sequence of events in the system call entry and exit, as illustrated in Figure~\ref{fig:hkpts}. A standard \texttt{read(fd, buf, count)} syscall passes through two of these eBPF hook points:

\begin{itemize}

    \item \textbf{Entry Hook (kprobe/sys\_enter):} As shown in the timeline (refer Figure~\ref{fig:bash_injection}), when the kernel receives the syscall, the arguments (file descriptor and buffer address) are visible. However, the \textit{User Buffer State} is currently \textbf{EMPTY} or stale because the kernel has not yet fetched data from the disk. The attack utilizes this stage solely to capture the target process ID and save the address of the user-space buffer (\texttt{buf}) into a BPF map.

    \item \textbf{Kernel Execution:} The kernel performs the internal work (e.g., reading the storage device) and populates the buffer with the legitimate script content.

    \item \textbf{Exit Hook (kretprobe/sys\_exit):} This is the vulnerability window. As indicated in Figure~\ref{fig:hkpts}, the \textit{User Buffer State} is now \textbf{FILLED WITH DATA}, but control has not yet returned to the User space process (Bash).
\end{itemize}

\begin{figure}[htbp]
	\centering
	\includegraphics[scale=0.6]{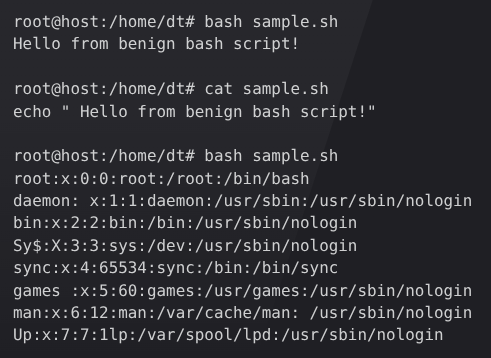}
	\caption{Output of the bash injection attack on the victim side}
	\label{fig:outbashinjection}
\end{figure}

The exploit triggers at this specific Exit point. The eBPF program retrieves the saved buffer address from the map and invokes \texttt{bpf\_probe\_write\_user}. Since the legitimate file content has already been written by the kernel, the malicious overwrite supersedes it. Finally, \texttt{bpf\_override\_return} is used to adjust the return value (byte count) to match the length of the injected malicious command, ensuring the Bash interpreter executes the payload cleanly without syntax errors.
Figure~\ref{fig:outbashinjection} illustrates the execution of an initially benign bash script. However, once the malicious eBPF program is loaded from the attacker's container, executing the same bash script results in the execution of the attacker's harmful payload instead of the original content.

\section{Possible Mitigations}\label{Sec:mitigations}

It is impossible to overlook the major attack surface that is introduced by the merging of cloud-native paradigms and 5G Service-Based Architecture (SBA). As demonstrated in our analysis, the shared kernel architecture of containerized deployments inherently exposes 5G Network Functions (NFs) to cross-container attacks via eBPF. Given that the aim of the modern 5G deployments is to improve modularity and performance, these vulnerabilities represent a systemic threat rather than isolated implementation bugs. We discuss several potential mitigation strategies and their feasibility below.

\subsection{Isolation via Virtualization}
One traditional approach to mitigate shared-kernel vulnerabilities is to replace containers with Virtual Machines (VMs). By providing each NF with its own guest kernel, the impact of a compromised eBPF program is limited to the VM instance, preventing direct access to the host or other NFs. However, this approach introduces significant resource overhead and latency. Furthermore, while VMs provide better isolation, they are not a perfect solution as they still remain vulnerable to the attacks that affect the host system itself. Moreover, the industry trend remains  skewed toward containerization.

\subsection{Capability and Feature Restriction}
The most direct defense against eBPF abuse is to restrict the privileges granted to the container. Removing the \texttt{CAP\_SYS\_ADMIN} and \texttt{CAP\_BPF} capabilities would effectively neutralize the attacks described in Section~\ref{Sec:vul_analysis}. However, this may be operationally infeasible in every scenario. In our case, these permissions are required by containers that manage the network interfaces or, in some cases, even to load eBPF programs. It is also required by the containers that use eBPF-based solutions for security and monitoring. Similarly, disabling the \texttt{bpf()} syscall entirely on the host would cripple the high-performance data paths required by 5G use-cases. Thus, a binary ``all-or-nothing'' permission model is insufficient.

\subsection{Fine-Grained Permission Models (LSM)}
Since 5G NFs require eBPF functionality, the defense strategy must shift from broad capability checks to fine-grained access control. A promising direction is the utilization of Linux Security Modules (LSMs), such as SELinux and AppArmor. A robust permission model should enforce:
\begin{itemize}
    \item \textbf{Hook Restriction:} Limiting which kernel hooks (e.g., \texttt{sys\_enter}, \texttt{kprobe}) a specific container can attach to.
    \item \textbf{Helper Function Allow-listing:} restricting access to dangerous helpers.
\end{itemize}

\subsection{Supply Chain Considerations}
It is crucial to recognize that even with a robust permission model, the system remains vulnerable to supply chain attacks. If a malicious eBPF program is embedded within a ``trusted'' vendor image that is built with necessary permissions, it may still abuse those rights. Therefore, mitigation efforts must extend beyond the scope of 5G networking protocols, and security must be addressed at the  system configuration level.

\section{Conclusion}\label{Sec:Conclusion}
	In this paper, we studied a vulnerability analysis of eBPF-enabled containerized 5G Core networks. By deploying a realistic testbed utilizing Open5GS and an eBPF-accelerated User Plane Function (UPF), we demonstrated that the shared kernel architecture inherent to cloud-native deployments exposes critical network functions to various threats. Our experimental results validated that a compromised container, possessing standard network privileges, can leverage eBPF to bypass namespace isolation. We successfully executed attacks ranging from passive reconnaissance and information theft to active DoS and privilege escalation via Bash injection. These findings highlight that while eBPF is indispensable for meeting 5G latency and throughput requirements, its integration requires a model that also handles the security implications that come with it.

	In our future work, we plan to expand this analysis to other prominent 5G software stacks, such as Free5GC~\cite{Free5GCOpenSource} and OpenAirInterface (OAI)~\cite{OpenairinterfaceOpenSource}, to evaluate the universality of these vulnerabilities across different implementation choices. Furthermore, we aim to design and implement a robust mitigation framework specifically tailored for 5G environments ensuring that high-performance packet processing can coexist with safety of the network.


\bibliographystyle{IEEEtran}
\bibliography{references_yash}
\end{document}